\documentclass[final,authoryear,1p,times]{elsarticle}

\usepackage{amssymb}

\journal{New Astronomy}

\begin{document}

\begin{frontmatter}

\title{Observational constraints on the nature of very short gamma-ray bursts}

\author{B. Czerny$^1$, A. Janiuk$^1$, D.B. Cline$^2$,  
S. Otwinowski$^2$}

\address{$^1$Nicolaus Copernicus Astronomical Center, Bartycka 18,
            00-716 Warsaw, Poland, e-mail: bcz@camk.edu.pl, agnes@camk.edu.pl \\
$^2$University of California Los Angeles, Department of Physics and 
Astronomy, Box 951447, Los Angeles, CA 90095-1547, USA, e-mail: stanislaw.otwinowski@cern.ch}

\begin{abstract}

We discuss a very peculiar subgroup of gamma-ray bursts among the BATSE 
sources. These bursts are very short ($T_{90} \le $0.1 s), hard, and came 
predominantly from a 
restricted direction of the sky (close to the Galactic anti-center).
We analyze their arrival times and 
possible correlations, as well as the profiles of individual bursts.
We find no peculiarities in the arrival times of Very Short Bursts (VSBs)
 despite their highly non-uniform
spatial distribution. There is no dependence in the burst shapes on location. Bursts
coming both from the burst-enhancement Galactic Anticenter region and from all other directions
show considerable dispersion in their rise and fall times. Significant fraction of VSBs have multiple peaks despite their extremely short duration. 
Burst time properties are most likely to be consistent with two origin mechanisms: either with
binary NS-NS mergers with low total masses passing through a phase of hypermassive neutron star,
or with evaporation of the primordial black holes in the scenario of no photosphere formation.
\end{abstract}

\begin{keyword}
gamma ray bursts \sep primordial black holes \sep BATSE

\end{keyword}

\end{frontmatter}

\section{Introduction}
\label{sect:Intro}

Observations of gamma ray bursts by BATSE satellite allowed to accumulate
the data set large enough for very detailed statistical studies. The 
distribution of the burst duration, defined as $T_{90}$ (time-interval during
which 90 per cent of the fluence is accumulated), shows clearly bi-modal
distribution \citep{Kouveliotou93,Norris2000}. 
Long bursts ($T_{90} > 2 $ s) show usually rather 
complex time-profiles, their
energy spectra are generally softer, and a number of these bursts are now
identified with the X-ray, optical and radio counterparts. Studies of the detected 
afterglows (see J. Greiner's web page\footnote{http://www.mpe.mpg.de/~jcg/grbgen.html}) 
clearly support the cosmological origin of these 
bursts \citep{Costa1997,vanParadijs1997,Metzger1997}.
The hypernova scenario, advocated strongly by \citet{Paczynski1995}, is the most 
plausible explanation of this phenomenon \citep[for a review, see][]{Piran2000,Mesarosz2002}.
 For short bursts, the most popular scenario is
compact object mergers \citep{Lattimer1976}; for a review see 
e.g. \citet{Bulik2004}.
 
Recently, \citet{Zhang2009} discussed the dichotomy of the two GRB populations and
introduced the nomenclature of Type I/II bursts, instead of the short/long bursts. 
This was
 to denote the 
physically distinct categories of cosmological GRBs and to stress that the
duration and hardness of the GRBs do not have to be the reliable indicators of their
true physical nature.

However, the existence of a separate third group of extremely short 
bursts of still another origin is 
also likely.  
\citet{Cline1999} suggested that very short bursts, 
lasting below 0.1 s, form a separate special subgroup of short bursts. They 
argued that the very short bursts have euclidean spatial distribution, rather 
different from the spatial distribution of bursts lasting 0.1 - 2 s 
\citep[as studied later e.g. by][]{Guetta2005}. They suggested
a local origin of these shortest bursts, possibly due to the evaporation of 
primordial black holes. The argument was further strengthened by \citet{Cline2003}, 
and by \citet{Cline2006} where they stress 
exceptional hardness of those events at the basis of KONUS data. 
These bursts were also found to
have clearly anisotropic distribution across the sky \citep{Clineproc2001}.

Careful analysis of the spatial distribution of short bursts ($T_{90} < 2 $ s)
was performed by \citet{Mag2003}. They showed an
enhanced correlation between the short bursts at angular scale of 2-4$^\circ$,
which might suggest that up to $\sim 13$ \% of the bursts may represent repeated
activity. The anisotropy in the burst distribution in a form of a spot around
Galactic coordinates l = 115$^\circ$ and b=30$^\circ$
 was argued for by \citet{Litvin2001}.

The properties of the short burst class are reviewed e.g. by  \citet{Nakar2007}.
The classical distinction between the short and long bursts is defined by the 
$T_{90}< 2$ s, and their duration distribution peaks at $T_{90} \approx 0.8 $ s.
The temporal structure of these bursts is  difficult for studies, because
of the low signal-to-noise ratio and limited time resolution.
Therefore, any more detailed studies of the temporal structure of the short bursts,
performed in the case of the long GRBs, were not carried out so far, because of 
the difficulty in analysis of their lightcurves.

\citet{Nakar2002} find that most of the bursts from their sample of 33 bright 
events exhibit a rapid time variability on timescales much shorter than their 
total durations. In their sample, more than a half of bursts show at least two well 
separated pulses, and more than a third show rapid variability. For the single pulses, 
their durations range from 5 to 300 ms, with a peak around 50 ms.

The pulse properties of 55 short bursts from BATSE were also studied by 
\citet{McBreen2001} who found that the rise and fall times as well as the FWHM and 
amplitudes show 
lognormal distributions. The median value of the rise time of a single pulse given by these authors
is 0.035 s, while for the fall time it is 0.056 and the asymmetry ratio is 
$t_{\rm rise}/t_{\rm fall}=0.65$. The median total duration time for this sample 
was 0.095 and the median number of pulses was 2.5. 

A possible sub-class of the short gamma ray bursts are the soft gamma-ray repeaters 
\citep[SGR:][]{Woods2006}. Their giant flares of durations about 0.1 s and hard 
($\sim 300$ keV) energy radiation are followed by a long (about 300 s), pulsating 
tail due to the rotation of a magnetized neutron star. The latter is much less 
luminous, and might easily be missed by the BATSE detector, in which case the event 
would be classified as a short GRB. 
\citet{Tanvir2005} found a correlation between the locations of short 
GRBs and positions of galaxies in the local Universe, indicating that 10-25\% of the events originate locally, at $z<0.025$. However, the signal was strongest for 
$T_{90} > 0.5$ s, which is longer than a typical duration of a SGR spike.
Furthermore, \citet{Ofek2007} finds, that the upper 
limit on the fraction of SGRs out of the BATSE sample is $f_{\rm SGR}<0.14$.

In the present paper we revisit the problem of the very short bursts (VSB), with
the aim to test their timelike characteristics. We use the BATSE sample
since the number of VSB with known localization on the sky in SWIFT data (only 8
events) is not high enough for our study.
In Section 
\ref{sec:batse_sample} we present our sample of very short bursts from BATSE.
We study their properties such as spatial distribution, arrival times and 
possible correlations, as well as the profiles of individual bursts.
In Section \ref{sec:diss} we discuss the possible origins of the special 
properties of our subsample and the most plausible mechanism of emission. 
In Section \ref{sec:concl} we conclude our work.

\section{Properties of the VSB from BATSE}
\label{sec:batse_sample}

Properties of the bursts detected by BATSE are provided by the public 
current catalog \footnote{http://http://f64.nsstc.nasa.gov/batse/grb/}. 
It contains
2702 events. Out of those, 2040 bursts have the measurements of 
$T_{90}$ as well as the coordinates in the sky.

The subsample of short bursts (hereafter SB), with $T_{90} < 2 $, consists
of 499 bursts. Out of those bursts, 51 have $T_{90} \le 0.1 $ s, and they
further form our subsample of VSB. As shown in Cline et al. (2003), 21 of those
come from 1/8 of the sky, extending from +90 to + 180 deg in galactic longitude and from -30 to + 30 deg in the galactic latitude (roughly in the direction of the Galactic Anticenter) so their spacial distribution is highly 
non-uniform. We now study other properties of this subsample 
which may make them different from
the usual bursts and indicate their nature. 

\subsection{Arrival times}

\begin{figure}
   \centering
\includegraphics[width=8cm]{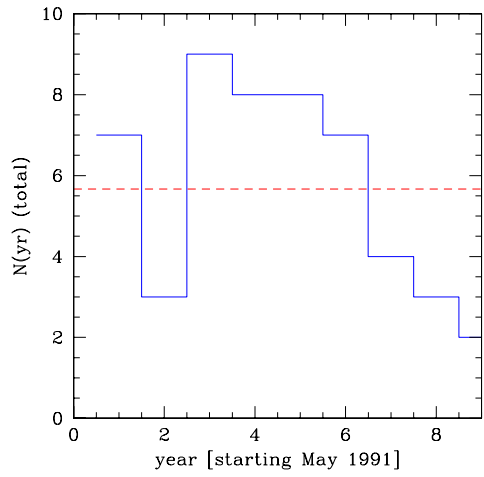}
\caption{Distribution of the bursts arrival times for VSB.}
\label{fig:times1}
\end{figure}

We construct a histogram of the arrival time distribution of VSB at yearly
basis. The result is shown in Fig.~\ref{fig:times1}. The data for all years 
were used, so in order to include as many bursts as possible we count the years 
from the beginning of the BATSE operation, i.e. the first year extends from
May 1991 till April 1992 etc. There seems to be a drop in the number of
VSB after 1997, and the distribution peaks around the years 1993 - 1995. 
However,  the probability that the
histogram describes the constant rate can be accepted at a 0.89 confidence 
level using the KS test although the straight line representing the mean rate 
(5.7/year) gives
a poor fit to this distribution ($\chi^2 = 13.5$ for 8 d.o.f.).

\begin{figure}
   \centering
\includegraphics[width=8cm]{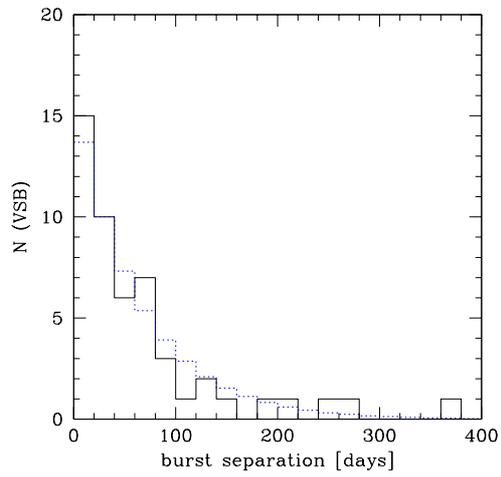}
\caption{Distribution of the bursts separation times for VSB. 
Expected distribution for random arrival times is marked with
the dotted histogram.}
\label{fig:times_differ}
\end{figure}

Any departure from a constant rate would be rather surprising effect.  
Since the bursts seem to be grouped, we also construct a histogram of the
time separations between the consecutive bursts. 
The result is 
shown in Fig.~\ref{fig:times_differ}. The expected average 
separation is
64 days. The observed distribution roughly corresponds to
the Poisson distribution of exponential decrease, as expected if the 
bursts are 
uncorrelated in time. However, traces of enhanced small separations as well as
big once may hint toward departure from truly random arrival.

We also attempted a study of the arrival times separately for the bursts located in the Galactic Anticenter region (VSB1) and bursts located outside this region (VSB2). The number of bursts inside the region is
enhanced at the beginning and seems to systematically decrease closer to the
end of observation. The time distribution of bursts outside the Anticenter region
seems to be more uniform. The $\chi^{2}$ test of hypothesis of both distributions
being constant in time give the fit quality of 11.0 and 13.0 per 8 d.o.f., for
VSB$_{1}$ and VSB$_{2}$, correspondingly.
The Kolmogorov-Smirnov test gives the
probability $P=0.567$ of the VSB$_{1}$ distribution and whole VSB sample 
being the same.
The distribution of the time separations for the whole VSB set and the bursts from
the Galactic Anticenter region 
was compared using the Kolmogorov-Smirnov test. We multiplied the
time separations for bursts from the Anticenter region by the ratio of the
total number of bursts in the region to the total number of VSB. After such
rescaling, the two distributions are consistent at the probability level of
0.976.

Therefore we conclude that the bursts which come from the restricted 
region in the Galaxy, roughly towards the Galactic Anticenter, and the other
VSB are not statistically different, and their arrival is statistically 
consistent with being uniform in time. 

\subsection{Burst profiles}

\subsubsection{Individual profiles}

\begin{figure*}
\includegraphics[width=18cm]{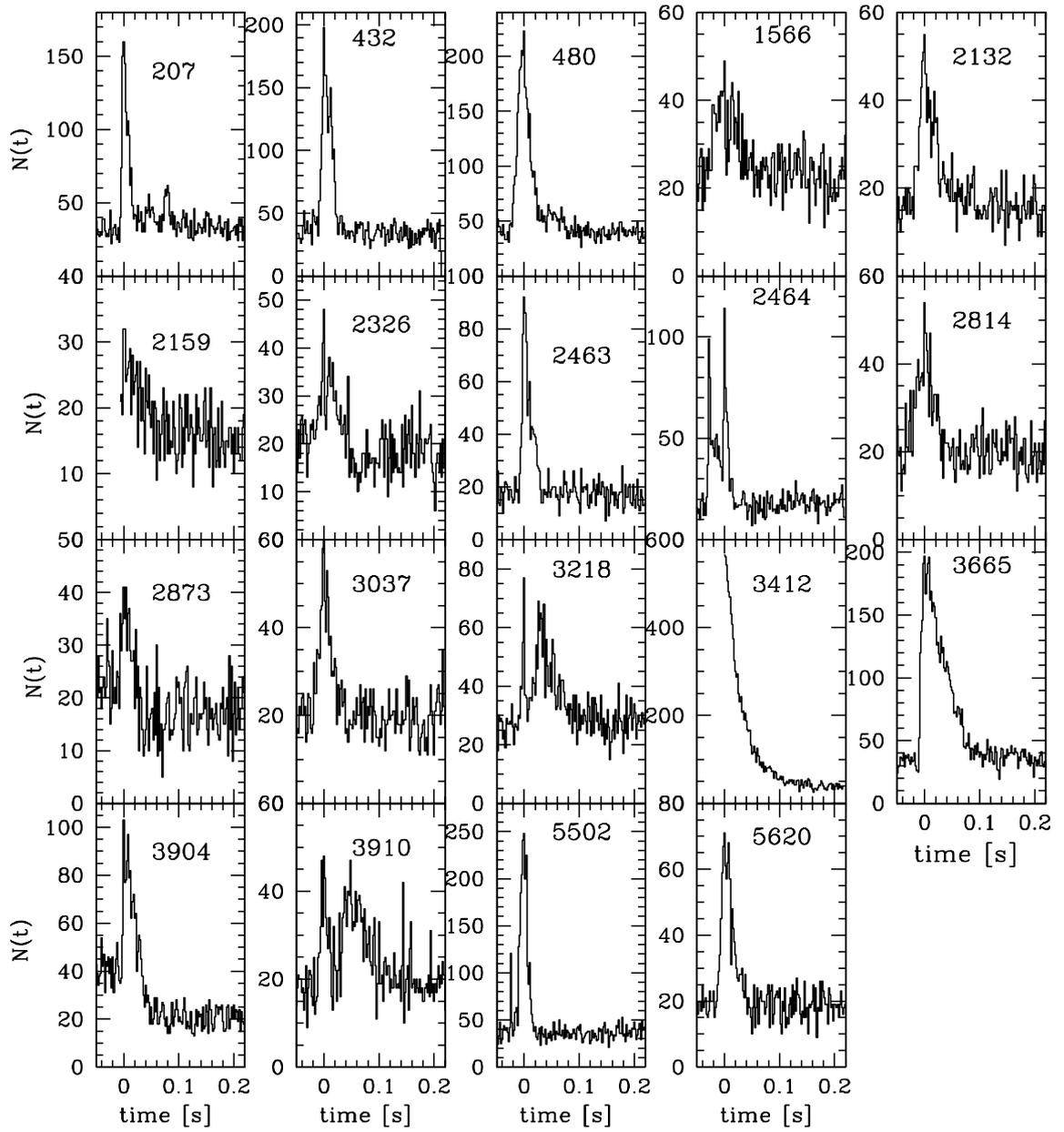}
\caption{Profiles of the VSB from the Galactic Anticenter region in channels 1 - 4.
 Count rate given for 
0.002 s bins.}
\label{fig:profile_all}
\end{figure*}

\begin{figure*}
\includegraphics[width=18cm]{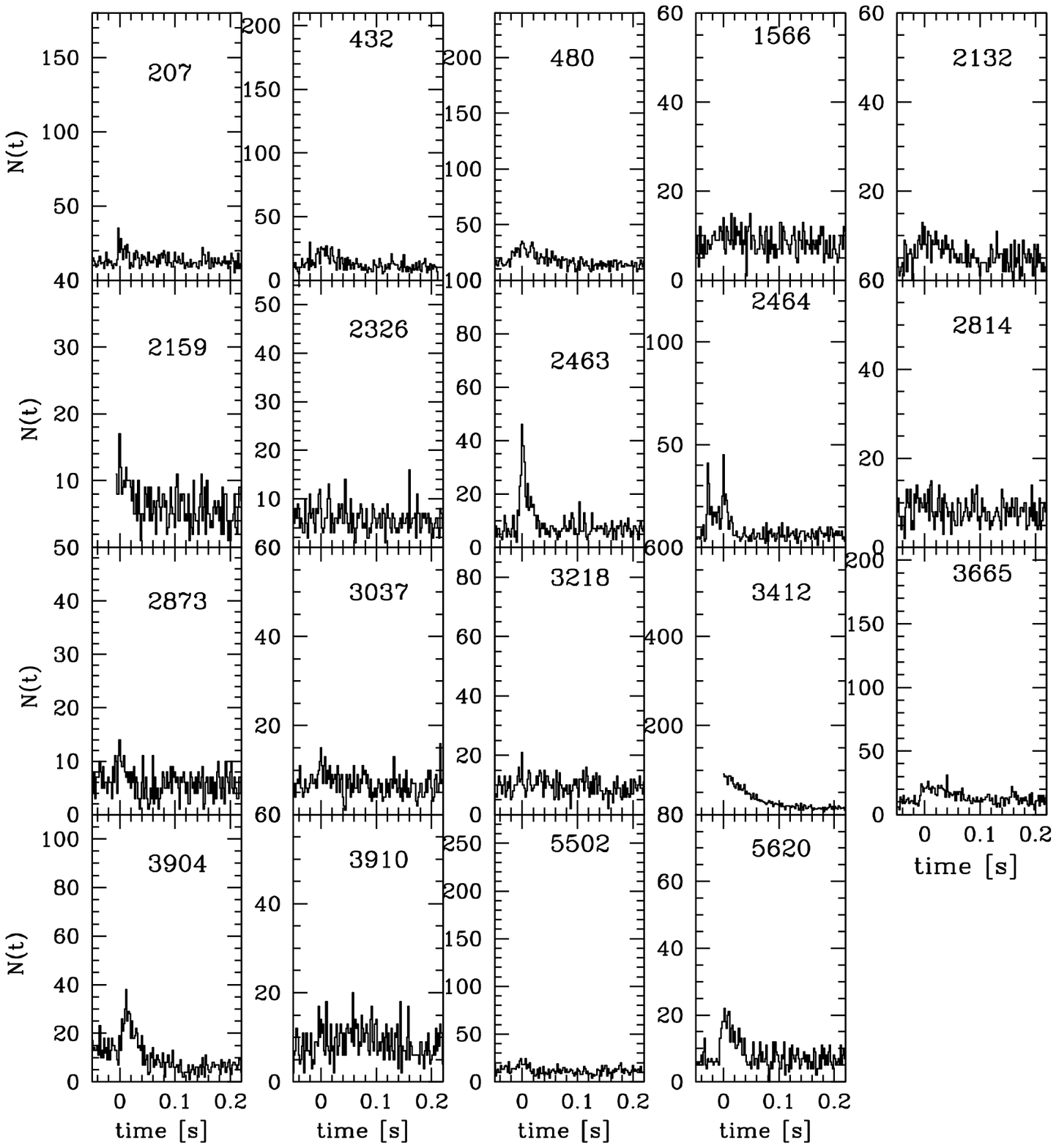}
\caption{Profiles of the VSB from the Galactic Anticenter region in channel 1.  Count rate given for 
0.002 s bins.}
\label{fig:profile_inside_1}
\end{figure*}

\begin{figure*}
\includegraphics[width=18cm]{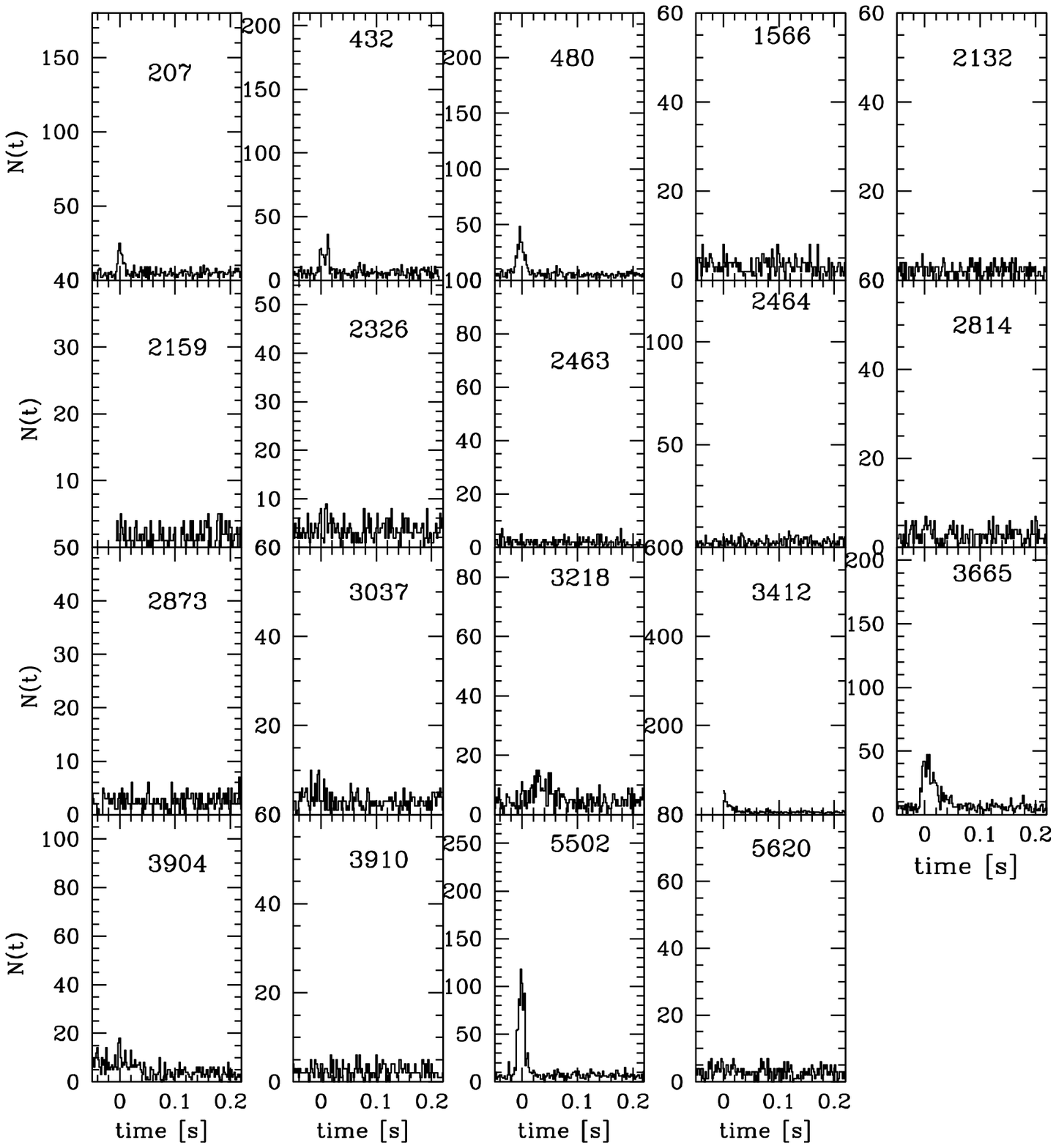}
\caption{Profiles of the VSB from the Galactic Anticenter region in channel 4. Count rate given for 
0.002 s bins.}
\label{fig:profile_inside_4}
\end{figure*}

\begin{figure*}
\includegraphics[width=18cm]{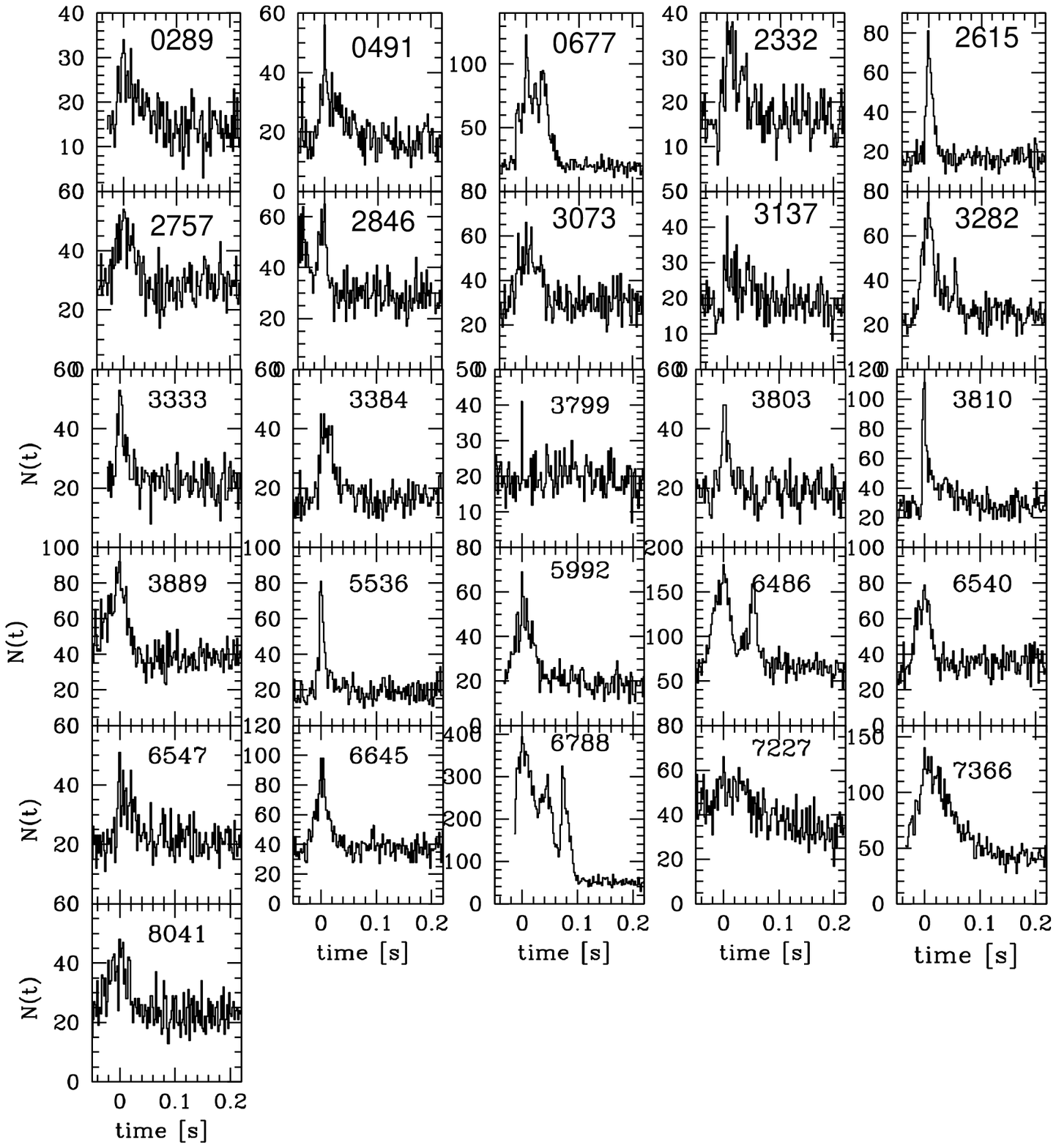}
\caption{Profiles of the VSB from outside the Galactic Anticenter region in channels 1 - 4. Count rate given for 
0.002 s bins.}
\label{fig:profile_all_outside}
\end{figure*}

\begin{figure*}
\includegraphics[width=18cm]{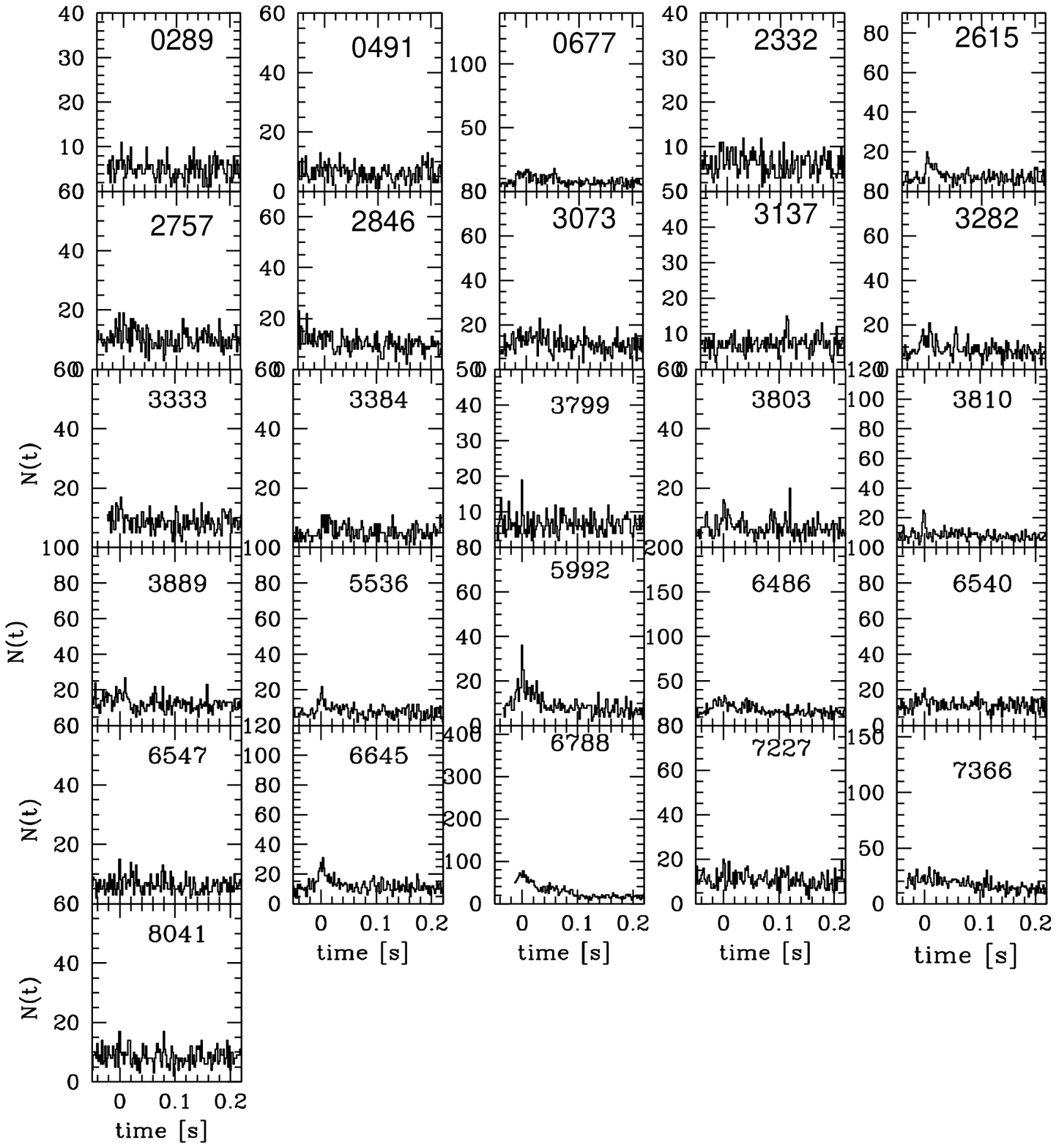}
\caption{Profiles of the VSB from outside the Galactic Anticenter region in channel 1. Count rate given for 
0.002 s bins.}
\label{fig:profile_all_outside_1}
\end{figure*}

\begin{figure*}
\includegraphics[width=18cm]{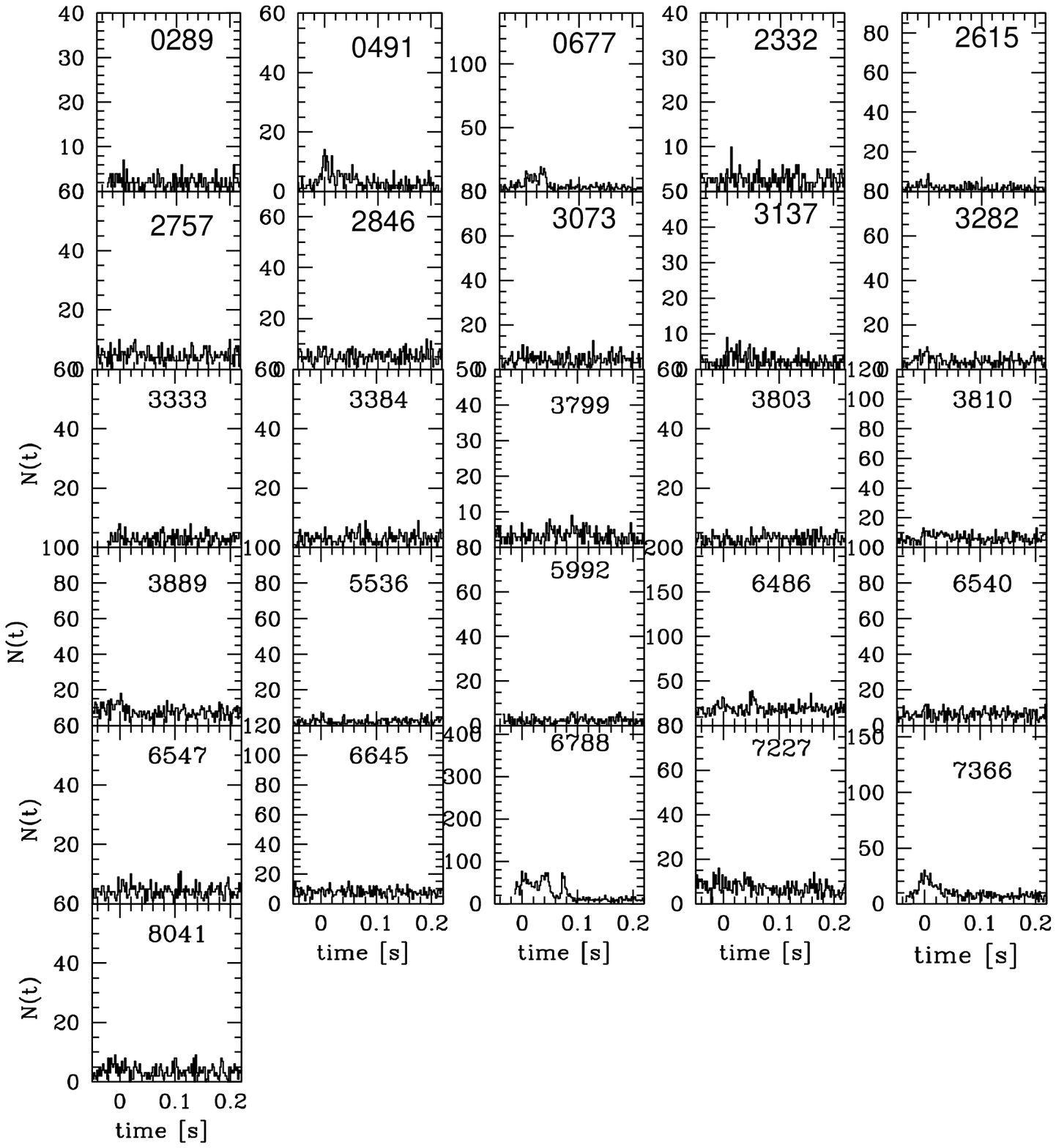}
\caption{Profiles of the VSB from outside the Galactic Anticenter region in channel 4. Count rate given for 
0.002 s bins.}
\label{fig:profile_all_outside_4}
\end{figure*}

We analyze the time profiles of the VSB from the event files 
for SB in the BATSE archive\footnote{http://cossc.gsfc.nasa.gov/batse/batseburst/tte/ascii.html}. 
The requested data for 44 bursts of all VSB are
available, including 19 bursts from the Anticenter region. We construct the 
lightcurve for each of the bursts using either all four channels, or
selected channels.  The individual profiles of bursts from the Galactic Anticenter region are shown in Fig.~\ref{fig:profile_all} (all four channels), Fig.~\ref{fig:profile_inside_1} (channel 1) and  Fig.~\ref{fig:profile_inside_4} (channel 4), and bursts from all other regions of the sky are shown in  Fig.~\ref{fig:profile_all_outside} (all four channels), Fig.~\ref{fig:profile_all_outside_1} (channel 1) and  Fig.~\ref{fig:profile_all_outside_4} (channel 4).  The bursts are labeled using their BATSE trigger number. 

All bursts show significant detections in the fourth channel. We determine the 
background in each channel separately and measure the peak countrate to the 
background ratio in 0.002 s bins. The ratio varies from  2.2 (for 6540) to 16.9 
(for 5502). All VSB are hard, as discussed before by Cline et al. (2005) at the 
basis of KONUS data.

Some of the bursts are clearly double-peaked (e.g. 2464, 3910, 6486, 6788),
for one of the bursts (3412) the {\sc tte} data starts after the peak, and for 
some bursts the data quality is so low that they cannot be used for the profile 
fitting analysis.

Therefore, we perform the profile fitting only for 17 bursts, using all four 
energy channels. Additionally, we
include the burst with the trigger number 0512 since it is very short
 although the formally given $T_{90}$ in BATSE catalog is somewhat higher than 0.1 s.

We subtract the
background for each of the bursts separately, and the background level is
determined from the data points separated by more than 0.15 s from the burst
maximum, and we renormalize the bursts to have maximum at 1.0 for all events.

\begin{figure*}
\includegraphics[width=16cm]{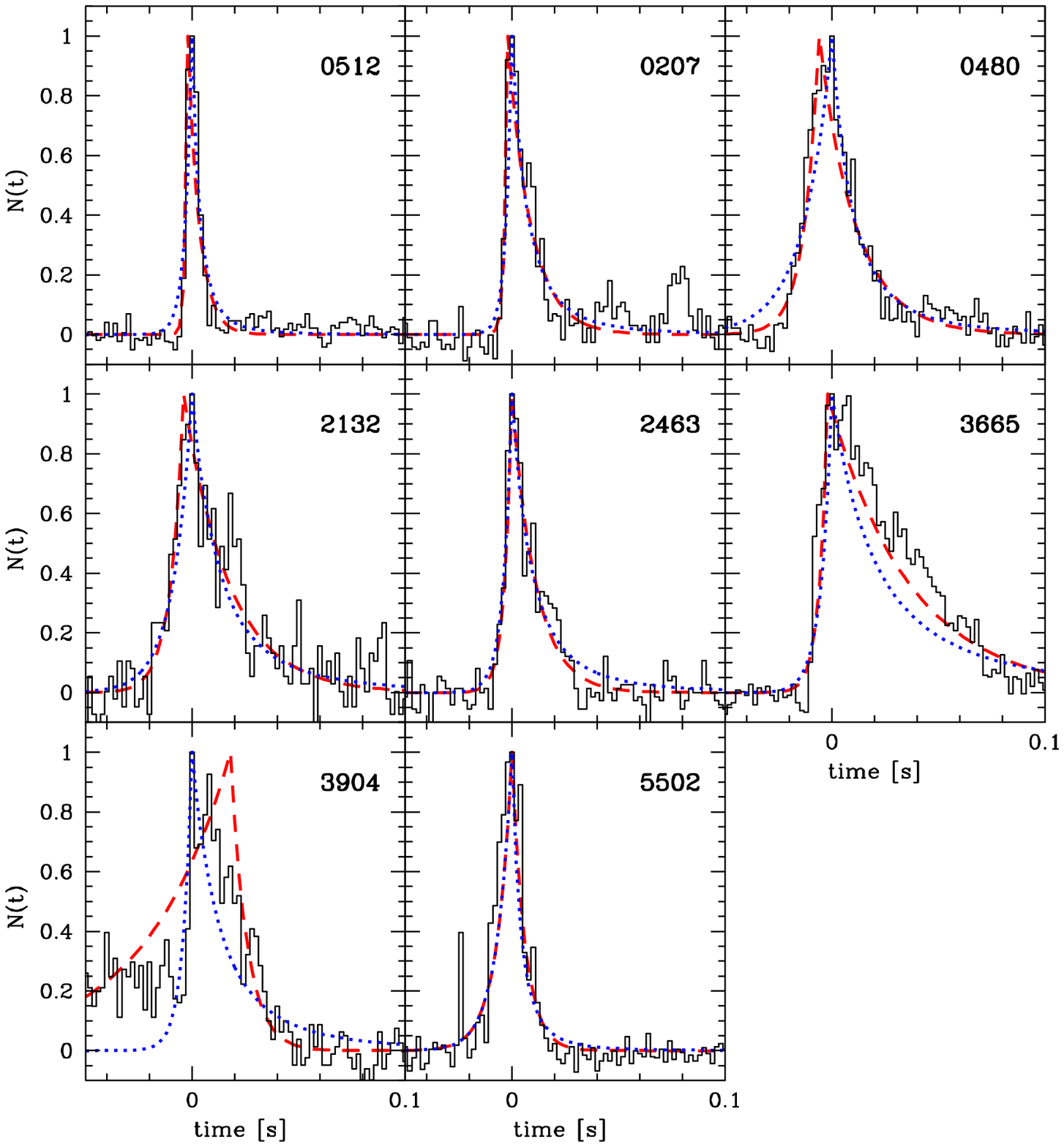}
\caption{Normalized background-subtracted profiles in 1 - 4 energy channels of the selected VSB 
from the Galactic Anticenter region in 
0.002 s bins. The short-dashed line shows the double exponential fit 
to the burst profile with fixed $t_{0}=0$, while the long-dashed line shows the fit with $t_{0}$ being a fitting parameter.}
\label{fig:indiv_inside}
\end{figure*}

\begin{figure*}
\includegraphics[width=16cm]{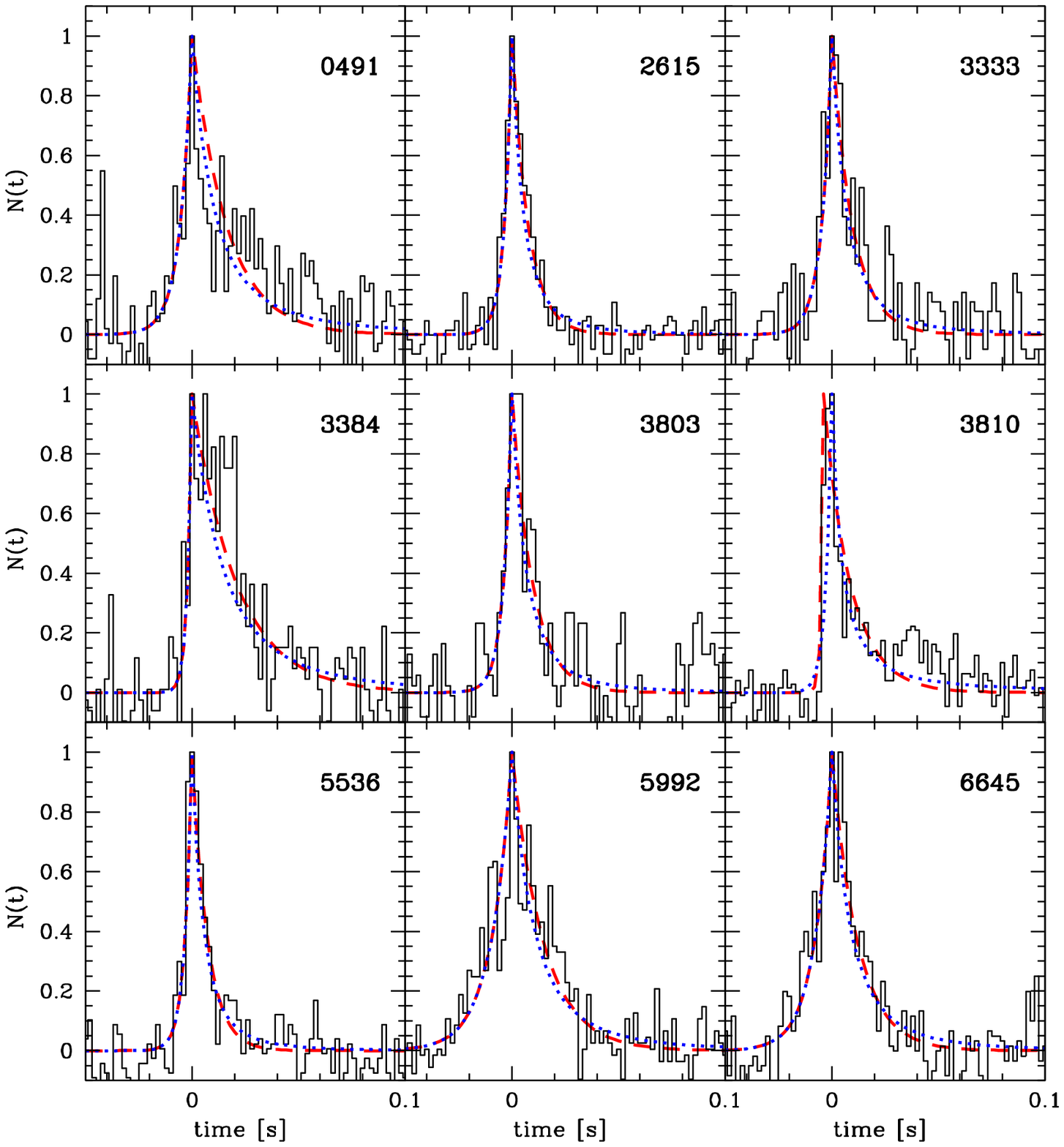}
\caption{Normalized background-subtracted profiles in 1 - 4 energy channels of the selected VSB 
from outside of the Galactic Anticenter region in 
0.002 s bins. The short-dashed line shows the double exponential fit 
to the burst profile with fixed $t_{0}=0$, while the long-dashed line shows the fit with $t_{0}$ being a fitting parameter.}
\label{fig:indiv_outside}
\end{figure*}

We fit the normalized profiles using double exponential fit:
\begin{eqnarray}
\label{eq:exponens}
f(t)& =& \exp((t - t_0)/t_{\rm rise})   ~~~~~~{\rm for ~t < t_0} \nonumber \\
    & = & \exp(-(t - t_0)/t_{\rm fall}) ~~~~~~{\rm for ~t > t_0} 
\end{eqnarray}
and the three parameters, $t_{\rm rise}$, $t_{\rm fall}$, and the shift $t_0$ are 
fitted to the data. The time axis for each burst is defined in such a way that $t = 0$ 
corresponds to the peak maximum (or to the first peak, if two equal count-rate peaks 
exist). We also made fits with $t_0 = 0$, i.e. with the change from rise to fall fixed 
at the maximum of the burst. However, better $\chi^2$ is in general given by fits 
with $t_0$ as a free parameter. In case of one of the bursts 
(BATSE trigger number 3904), a freedom of $t_0$ shifts the peak considerably and the 
fit is problematic due to the fact that the background before the burst is much 
higher in this case than after the burst. Both fits are presented: selected bursts, 
normalized and background-subtracted, are shown in Figs.~\ref{fig:indiv_inside} 
and \ref{fig:indiv_outside}, for bursts from the Galactic Anticenter region and for 
those from the other parts of the sky, respectively. In the 
further analysis we use the fit with arbitrary $t_0$ for all bursts for consistency. 

\begin{table*}
\caption{Properties of single VSB with good S/N ratio}
\begin{tabular}{lrrrr}
\hline 
\hline
BATSE trigger & $t_{rise}$ & $t_{fall}$ & $t_0$ & $\chi^2$/dof\\
              & [s]  & [s]  & [s]   \\
0512  & $0.0011_{-0.0004}^{+0.0003}$ & 0.0053 $\pm$ 0.0005 & -0.002 $\pm$ 0.002 & 1.49\\
0207  & $0.0014_{-0.0005}^{+0.0006}$ & 0.0108 $\pm$ 0.0015 & -0.002 $\pm$ 0.002 & 1.27\\
0480  & $0.0068_{-0.0009}^{+0.0009}$ & 0.0172 $\pm$ 0.0014 & -0.006 $\pm$ 0.002 & 1.35\\
2132  & $0.0061_{-0.0022}^{+0.0023}$ & 0.0212 $\pm$ 0.0041 & -0.004 $\pm$ 0.002 & 1.03\\
2463  & $0.0035_{-0.0009}^{+0.0010}$ & 0.0110 $\pm$ 0.0018 &    0.0 $\pm$ 0.002 & 1.31\\
3665  & $0.0038_{-0.0007}^{+0.0006}$ & 0.0379 $\pm$ 0.0007 & -0.002 $\pm$ 0.002 & 1.72\\
3904  & $0.0397_{-0.0048}^{+0.0050}$ & 0.0076 $\pm$ 0.0017 &  0.018 $\pm$ 0.002 & 2.17\\
5502  & $0.0071_{-0.0008}^{+0.0007}$ & 0.0060 $\pm$ 0.0008 &    0.0 $\pm$ 0.002 & 2.26\\
0491  & $0.0056_{-0.0022}^{+0.0023}$ & 0.0161 $\pm$ 0.0046 &    0.0 $\pm$ 0.004 & 1.28\\
2615  & $0.0036_{-0.0011}^{+0.0011}$ & 0.0076 $\pm$ 0.0017 &    0.0 $\pm$ 0.002 & 1.06\\
3333  & $0.0048_{-0.0024}^{+0.0030}$ & 0.0097 $\pm$ 0.0038 &    0.0 $\pm$ 0.003 & 1.12\\
3384  & $0.0021_{-0.0021}^{+0.0022}$ & 0.0207 $\pm$ 0.0049 &    0.0 $\pm$ 0.002 & 1.01\\
3803  & $0.0040_{-0.0023}^{+0.0025}$ & 0.0096 $\pm$ 0.0036 &    0.0 $\pm$ 0.003 & 1.38\\
3810  & $0.0010_{-0.0010}^{+0.0003}$ & 0.0126 $\pm$ 0.0025 & -0.004 $\pm$ 0.002 & 1.29\\
5536  & $0.0034_{-0.0012}^{+0.0011}$ & 0.0072 $\pm$ 0.0019 &    0.0 $\pm$ 0.002 & 1.17\\
5992  & $0.0098_{-0.0027}^{+0.0026}$ & 0.0146 $\pm$ 0.0031 &    0.0 $\pm$ 0.003 & 1.24\\
6645  & $0.0080_{-0.0022}^{+0.0024}$ & 0.0120 $\pm$ 0.0028 &    0.0 $\pm$ 0.002 & 1.00\\
\hline
\end{tabular}
\label{tab:expfits}
\end{table*}

\begin{figure}
\includegraphics[width=8cm]{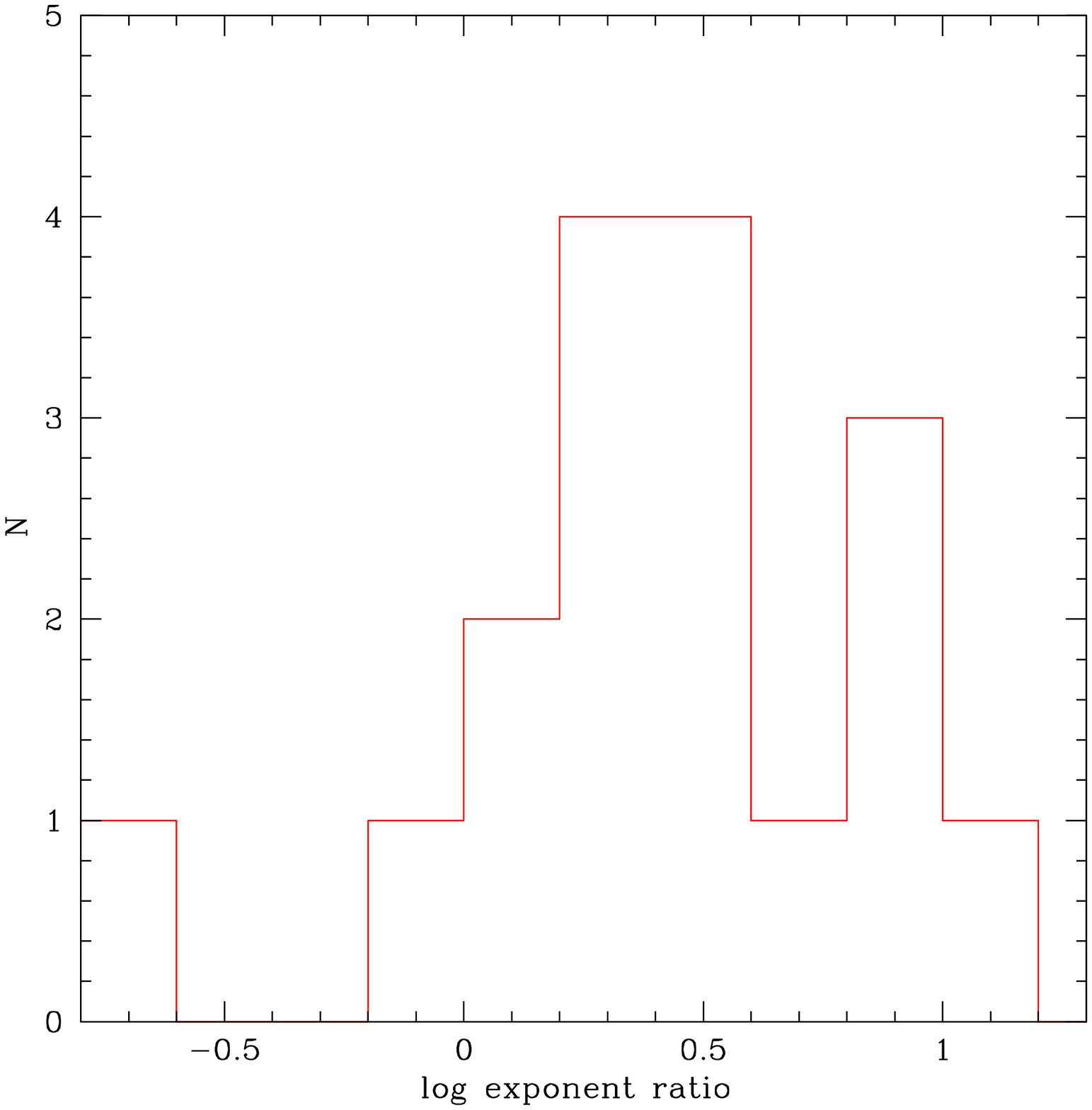}
\caption{Histogram of the exponent ratio $r = t_{\rm fall}/t_{\rm rise}$ with 
$t_{\rm rise}$ and $t_{\rm fall}$ defined on the rising and falling part of the 
lightcurve of selected VSB}
\label{fig:exp_rat}
\end{figure}

Almost all bursts show certain amount of 
asymmetry, with
faster rise time and slower decrease. 
The median of the rise timescale is 0.0040 s, the median of the fall timescale is 
longer, 0.0110 s, and the median value of the ratio of these two timescales is 0.4. 
Therefore, the timescales on average are much shorter than found by 
\citet{McBreen2001} for short bursts. This is because they included also short 
bursts with $T_{90} > 0.1$ s. The timescale ratio indicates larger asymmetry in 
selected 
VSB but this effect might be due to some contribution of double-peak events even in 
the selected bursts. For example, 2132 seems to have traces of a secondary maximum 
at $ t \approx 0.015$ s, and 3665 may actually be a multiple case.
In Fig.~\ref{fig:exp_rat} we show the distribution of the ratios of the two 
exponents in the
two-exponent fit to the selected bursts  (Equation~\ref{eq:exponens}). 
The histogram has two peaks, one around 
 $r=t_{\rm fall}/t_{\rm rise}$=2.5 ratio and one around the ratio of $r \approx$ 
10. However, the effect of bimodality is not significant, taking into account the small 
number of bursts. 

We checked the burst profiles in all four energy channels for possible time delays. In some
bursts the peak in the first channel was earlier than in the 
remaining three channels
(a negative lag in trigger no. 0207 by 2 ms; in 0289 by 4 ms; in 0491 by 8 ms; in 2326 
by 2 ms; in 3910 by 4 ms), and in a few cases the peak in the channel 4th was delayed 
(a positive lag in trigger no. 0432 by 12 ms; in 2159 by 2 ms; in 2332 by 8 ms; in 
3384 by 8 ms; in 3803 by 6 ms; in 5536 by 4 ms). 
In other bursts, the peak in channel 4 was 
advanced (a negative lag in trigger no. 2463 by 2 ms;  in 3037 by 4 ms; in 3810 by 
2 ms;in 5502 by 2 ms; in 5620 by 4 ms; in 6547 by 2 ms;
in 7227 by 2 ms). In two bursts (2757 and 3073) two peaks in channel 4 are seen. 
In all other bursts the 
peaks in all energy channels coincided (with the accuracy of the 2 ms time bin) 
within the 
statistical error of one sigma. However 
the poor countrate prevents us from the detailed 
fitting of the 
profiles in separate energy channels and the quantitative measurement of the time 
delays was not possible, and the lags mentioned above are likely insignificant. 

 We also compared the fluxes of the bursts within the Anticenter region and outside it.
The bursts seen towards the Anticenter are somewhat brighter (with median countrate of $77 \pm 5$ 
cts in 0.002 bins without background subtraction) than the remaining bursts (median of $66 \pm 6$ cts),
although the background itself determined close to VSB is actually slightly lower in the
direction of the Anticenter \citep[25.0 cts vs. 27.6 cts in 0.002 bins]{Cline2005}.
However, since errors are large the difference is not highly significant.

\subsubsection{Composite profiles}

In order to overcome the problem of low S/N ratio in most of the bursts
we also construct the composite burst profile. 

\begin{figure}
\includegraphics[width=8cm]{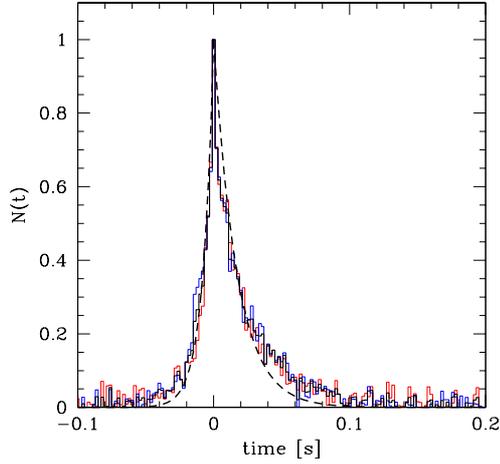}
\caption{Composite burst profiles for all VSB (black line), for bursts from the 
Galactic Anticenter region (red line) and for bursts from outside that region 
(blue line). The analytical fit (dashed line) is given 
by Eq.~\ref{eq:exponens}.
Better fit for the decay part is provided by 
Ryde \& Svensson function (Eq. \ref{eq:pl}; see text)}
\label{fig:composite}
\end{figure}

The background-subtracted lightcurve is later normalized by its
maximum value and shifted, so the maximum is at $t=0$. All lightcurves
in all four channels are added, and the result divided by the number of lightcurves
included. The resulting composite profile is shown in Fig.~\ref{fig:composite}.

The profile shows significant asymmetry, characteristic also for 
longer bursts. Its shape is well represented by the exponential rise with
the timescale of 0.0079 s,
and approximately represented by the exponential decay time of 0.0171 s.

These timescales are much shorter than the average rise and decay timescales of
single pulses in a sample of SB \citep[$\sim 0.036$ s and $ \sim 5.5$ s
correspondingly,][]{McBreen2001} but fall within the broad distribution
of these parameters in the sample. Therefore,
VSB show basically the same trend as seen in all gamma ray bursts 
\citep{Band1997,Norris2000} although the timescales involved are 
clearly much shorter.

Better fit to the decay phase, however, is provided by the function
\begin{equation}
f(t) = {1 \over ({1 + t/\tau })^{n}}
\label{eq:pl}
\end{equation}
with index $n = 2.62$ and the decay timescale $\tau = 0.0319$ s.
\citet{RS2002} analyzed 22 bright pulses in long bursts 
using this description and they found a bi-modal distribution. 
Although most of bursts are well represented by $n = 1.0$,
there was a secondary enhancement around $n = 3.0$. It is therefore interesting
to note that VSB may show some similarity to the latter subclass of pulses. 

If we choose only the energy range 50 - 300 keV (i.e. the channels 2+3) for
the analysis the results are: timescale of the 
exponential rise 0.0093 s, exponential decay 0.0195 s, 
with again a much better fit to the decay part given by Eq.~\ref{eq:pl}, 
with $n=2.44$ and $\tau = 0.0319$ s.  

We searched for a correlation between the burst asymmetry and the burst hardness defined
as a ratio of the peak countrate in the channel 4 to the total peak countrate.
However, 
we do not detect a significant trend, perhaps due to the low quality of the data.

\begin{figure}
\includegraphics[width=8cm]{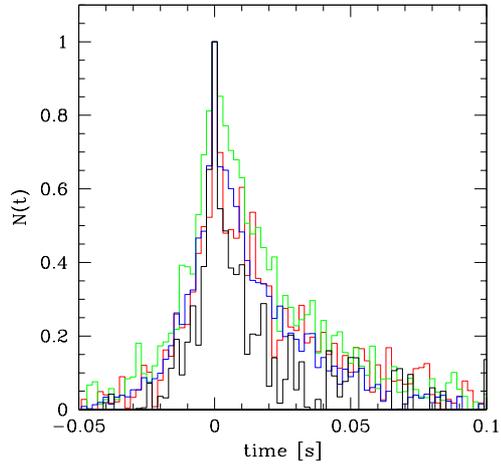}
\caption{Burst composite profiles for all VSB, for four energy channels separately, normalized to 1 for each energy channel: 
Red:channel 1, Green:channel 2, Blue:channel 3, Black:channel 4.
The narrowest profile is seen in channel 4.}
\label{fig:VSB_energy2}
\end{figure}

We also calculated the composite burst profile
in all four channels separately. We used the fixed time grid determined for 
the total
profile (e.g. for combined 1 - 4 channels). In this way we allowed for advancement or a delay of the burst
maximum with respect to the position of the maximum in the average profile
from all four channels. The result is shown in
Fig.~\ref{fig:VSB_energy2}.
Burst peaks still coincide within the energy grid accuracy (0.002 s).
 We cannot increase the time resolution since the number of available 
photons is not large enough. The graph shows, however, that the profile in
4th energy channel is the narrowest.

\begin{figure}
\includegraphics[width=8cm]{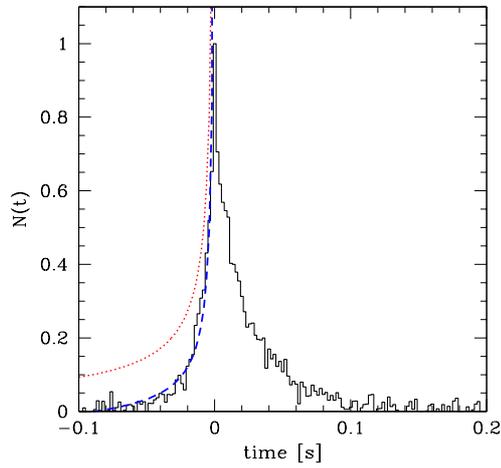}
\caption{Composite burst profile for all VSB (black line) with
the analytical fits given 
by Eq.~\ref{eq:hawking}. The dotted line shows the fit with normalization
calculated from the luminosity integral, while the dashed line
shows the profile with normalization fitted to the background level.}
\label{fig:hawking_fit}
\end{figure}

We also compare the shape of the composite pulse profile to specific prediction
of the model of evaporation of a primordial black hole 
\citep[for basic reviews, see e.g.][]{Carr1976,Carr2005}.

The time profile of the pulse produced due to the black hole evaporation can be 
derived from the temperature of the black body radiation, which in 
the Hawking process is inversely dependent on the black hole mass:
\begin{equation}
T = {\hbar c^{3} \over 8 \pi G k M} = 10^{16} {1 \over M[g]} [MeV].
\end{equation}
The luminosity is then given by:
\begin{equation}
L = - {dE \over dt} = {d (Mc^{2}) \over dt} = - \sigma T^{4} 4\pi R^{2},
\label{eq:lum1}
\end{equation}
where $R=2 G M/c^{2}$ is the Schwarzschild radius. From the above relation, 
one can see that the luminosity is inversely proportional to the square of the 
black hole mass, $M^{2}$. The mass dependence on time is found by integrating $dM/dt$
from the time $t<0$ to $t_{0}=0$, when the black hole is completely evaporated. The 
lifetime of a black hole till evaporation is
\begin{equation}
t \propto M^3,
\end{equation}
and the luminosity from Eq.~\ref{eq:lum1} can be expressed as a function of time:
\begin{equation}
L(t) \propto  C t^{-2/3}.
\label{eq:hawking}
\end{equation} 
The luminosity profile goes to infinity at $t=0$. Nevertheless, the integrated 
luminosity is
finite, so we can derive the normalization of the lightcurve
based on the integral before the burst maximum being equal to that after the maximum.

The fit of the function given by Eq. \ref{eq:hawking} 
to the composite profile is shown in Figure \ref{fig:hawking_fit}.
Two curves represent the fits with (i) the normalization derived form the integrated 
luminosity (i.e. the integral of the analytic curve and the background-subtracted curve 
coincide) and (ii) the normalization arbitrarily fitted to the
background level. The first method overpredicts significantly the luminosity of 
the burst
before the peak. However, since the evaporation formula above does not give the time 
profile after the peak, the back ground determination may be biased by enhanced emission 
before as well as after the peak. This argument justifies the second approach, and in 
this case the fit to the rising part of the profile is good. 

The approach used above neglects the change of the spectral shape in time.
It can be justified in the following way. The temperature of the black body 
emission during the final stage ($\sim 0.1$ s before complete evaporation) is
of the order of 15 TeV, and at that stage a fireball is already formed.
In other words, the
evaporating plasma (photons, electron-positron pairs and heavier particles)
is no longer optically thin, because a simple estimate of the total thickness for 
electron scattering indicates a depth of the order of $1000$.
The following creation of electron-positron pairs is efficient 
 and the final temperature 
of the radiation reaching an observer saturates at the energy treshold 
for this process.
If this approximation is indeed justified then there is no temperature 
variation with time at the final stages and only the luminosity varies.
Therefore the
specific energy-dependent sensitivity of an instrument does not influence the 
shape of the lightcurve. 

Also, if we assume that indeed the emitted particles do not interact, the
peak emission of photons is at very high energies, well above the instrument 
response. The detected photons would then come from the low energy tail,
again perhaps weakly dependent on time apart from normalization.

However, if the thermalization is not full, and for example pion
processes are important as
discussed in several  papers \citep[e.g.][]{Maki1996,Kapusta99,Carr2010}, or if
the relativistic boosting of the fireball varies with time, there may be a 
spectral 
hardening effect
and in this case our approach is too simplistic. However, no specific models
of time-dependent spectral evolution of a fireball are available yet. 

\section{Discussion}
\label{sec:diss}

Very Short Gamma-ray Bursts were proposed to originate as a result of evaporation of 
the primordial black holes \citep{Cline1999}. In this paper we carefully studied 
the spectral and temporal properties of VSB from BATSE with the aim to test whether 
they are likely to have 
a different origin than the other short gamma ray bursts. 

The burst duration for VSB is much shorter than the duration of short bursts, or of 
mixed samples with VSB and short bursts included \citep[e.g.][]{McBreen2001}. The 
shortest rise time is only 5.3 ms. However, otherwise they seem to be
the scaled-down analogues
of short bursts. They are typically asymmetric, with faster rise than the 
decay time. Moreover, a number of VSB still consist of two or more subpulses, so the 
complex structure typical for longer bursts is visible even in this class.

Significant fraction of VSB (21 out of 51) come from 1/8th of the sky located roughly 
in the direction of the Galactic Anticenter region, extending from +90 to + 180 deg 
in galactic longitude and from -30 to + 30 deg in the galactic latitude 
\citep{Cline2003}. 
The significance of this anisotropy has been estimated by \citep{Clineproc2001}
 to be $1.4 \times 10^{-6}$, and
by Piotrowski et al.\footnote{http://www.lip.pt/events/2006/ecrs/proc/ecrs06-s5-10.pdf},
(unpublished) to be a few times $10^{-5}$ using the factorial analysis. 
Bursts coming from the Anticenter region are slightly 
brighter but the effect is at the level of 2 sigma, and the 
distribution of the arrival times of both classes is statistically consistent 
with uniform. 

The enhancement of the number of VSB BATSE bursts in the direction of the 
Galactic Anticenter is clearly a puzzle.
On the other hand, there are several interesting observations indicating
the exceptional properties of that region. Roughly at that direction there is
one of the richest known regions of star formation in the Galaxy, with over 800
HII regions, many Wolf-Rayet stars, including the richest Cygnus OB2 
association \citep[see][for Cygnus-X Spicer Legacy Survey]{Kraemer2010}. 
The COMPTEL Al 26 survey indicated a clear enhancement in the Anticenter region
also likely related to nucleosynthesis activity 
\citep[see e.g.][]{Diehl2001}. An excess of the TeV cosmic rays from the general 
direction of the heliotail, also close to the Galactic Anticenter, has been 
reported by \citet{abdo2008}. However,  \citep{salvati2008} argued against the 
heliotail origin. Similar claims of such cosmic ray anisotropy has also been 
done in the past, but with the argument in favor of the connection with heliotail 
based on annual trend \citep{Fujimoto2001}. Interesting possibility of correlations 
between the VSB and the regions of enhanced correlation between the microwave 
background and the diffuse gamma rays above 1 GeV has been considered by 
\citet{Jedrzejczak2007}.

For the energy distribution, 
VSB are much harder than short bursts \citep{Cline2005}.
However, this effect 
can be considered as a continuation of a more general trend of bursts being harder 
when shorter, including the long-soft GRBs \citep[see e.g.][]{Zhang2009}. 
We did not find any significant time delays between the 
peak emission in the four energy channels and we can only put an upper limit
of 2 ms for such a delay. It is interesting to note that studies of long bursts 
showed that there is an anticorrelation
between the lags and the total burst isotropic luminosity \citep{Norris2000}:  
\begin{equation}
L_{\rm pk} \approx 1.3 ({\tau \over 0.01 s})^{-1.15} ~~[10^{53} ~erg ~s^{-1}]
\end{equation}
It was interpreted by \citet{Salmonson2002} as due to the
variation in jet opening angles. They also suggested that measurements of 
variability or of spectral lags can then be used as a crude distance indicator 
for the GRBs. However, in the case of short bursts, both positive and negative lags 
were reported \citep{Gupta2002}, usually on the order of 50 ms. Therefore, 
the absence of measurable lags in combined profiles of VSB may not be surprising.

The distribution of the GRBs in space shows that the cosmological effects
are playing the important role. The mean $V/V_{\rm max}$ value for the sample
of long bursts was 0.282, while for the short GRBs it was somewhat larger,
equal to 0.390 \citep{Guetta2005, Guetta2006}. In this respect, the value estimated by
\citet{Cline1999} to be 0.52, and indicating the euclidean distribution
of the VSBs, seems to be a continuation of this trend. The shorter the 
bursts, the more locally they are observed. 
Based on our present study, is still not clear whether this might be 
the effect of a yet another different progenitor type, or rather a selection effect.

\subsection{VSB from SWIFT}
\label{sect:swift}
 
The number of very short events ($T_{90} < $ 0.1 s) detected by the SWIFT satellite 
is small so they are not suitable
for statistical analysis. Nevertheless, these high quality data may shed some light
onto the nature of such events. In Table~\ref{tab:SWIFT} we list all the events with
$T_{90} < $ 0.1 s, and in addition two more with $T_{90} $ just above 0.1. Out of those 10 sources, none
is located in the Anticenter region found overpopulated with VSB in the BATSE data. 
Three bursts now populate another of the 8 sky regions discussed by \citet{Clineproc2001}, 
located between 0 and 90 deg of the Galactic longitude, and -30 and +30 deg of the Galactic latitude.
The probability of such coincidence is 0.09  which cannot be considered
as highly significant departure from uniform distribution.

Candidate afterglow/host galaxies were claimed for some of those bursts but
 fading of the optical
counterpart was never detected for any of those sources. For 100206A an optical counterpart 
was found by \citet{Levan2010}. The spectroscopy 
obtained with the Keck telescope revealed the presence of 
a galaxy at the indicated location, with the redshift 0.41 \citep{Cenko2010}, and the projected 
offset of the burst from the galaxy center was $\sim 35$ kpc. The burst GRB 060502B has been found
to coincide with $z = 0.287$ galaxy, and the implied offset from the center was $73 \pm 19$ kpc
\citep{Bloom2007}. The event 050509B was very bright in X-rays and well localized, but again 
no fading optical/near infrared/millimeter counterpart was found. The source has been tentatively
connected with $z = 0.225$ galaxy, a member of a galaxy cluster \citep{Gehrels2005,Hjorth2005,Castro-Tirado}. 
The projection distance in this case is $\sim 33 $ kpc. However, the X-ray afterglow measured by Swift 
XRT in this source starting from 62 s after the gamma burst was detected up to 300 s after the burst
\citep{Rol2005,Bloom2005}, and the X-ray luminosity was decreasing exponentially, with an index of $\sim 1.3$,
implying the decay time below $60$ s.
The relatively long duration of this signal in comparison with the prompt gamma-ray emission may 
indeed support some (relativistic?) ejection of material which later undergoes cooling. \citet{lee2005}
found this timescale roughly consistent with the binary merger scenario.

The comprehensive information on each of the bursts can be found through the internet
\footnote{http://www.swift.ac.uk},
\footnote{http://gcn.gsfc.nasa.gov}
 for each of the bursts. 
The burst profiles viewed in 2 ms bins used for BATSE data analysis are usually single,
although some substructure may be present in GRB 050925, GRB 070810B  and GRB 090417A. 

\begin{table*}
\caption{Properties of VSB from Swift}
\begin{tabular}{lrrrrr}
\hline 
\hline
name     & $t_{90}$ & gal. longitude & galactic latitude & redshift \\
         & [s]     &          &    \\
\hline
050509B   & 0.073  & 182.90828684   &   86.15594183 &  0.225 ?    \\  
050925   & 0.070  & 72.30514279    &  -0.07005621  &  -       \\
051105A  & 0.093$^a$  & 59.86196122  &    28.78790172 & -   \\    
060502B  & 0.131  & 81.59665275    &  23.53577034  & 0.287 ?\\
070209   & 0.090  & 259.76518949   &  -57.02797996    &    -      \\              
070810B  & 0.080  & 116.37637070   &  -53.85090173  &  -       \\   
070923   & 0.05 & 295.85521498 &    24.12470763 & -      \\  
090417A  & 0.072 & 173.46586705   &   -61.03545070 & -      \\
090515   & 0.036  &  232.46491111    &  60.44320773 & -  \\   
100206A  & 0.12  & 166.88023385  &   -37.72547634  &   0.41 ?  \\ 
\hline
\end{tabular}
\label{tab:SWIFT}
\\
$^a$ $T_{90} = 0.028$ according to \citep{Cummings2005}, other burst parameters according to
on-line SWIFT catalog \verb+http://heasarc.gsfc.nasa.gov/docs/swift/archive/grb_table/index.php+.
\end{table*}

\subsection{Possible mechanism of emission}

The tentative continuity of the burst properties from long bursts through short 
to VSB 
does not imply the same mechanism behind those events: long bursts are due 
to hypernova phenomenon, 
short burst are due to mergers of compact objects in binary systems (neutron stars 
or black holes), and the origin of the VSB is still waiting for explanation. Two 
possibilities are natural to consider: a binary merger but with (for some reason)
exceptionally short
timescale and evaporation of primordial black holes. We consider those 
two scenarios below.

\subsubsection{binary merger}

The merger of binary systems NS-NS and NS-BH were modeled by a number of authors 
\citep[for the review, see e.g.][]{Faber2009}. The expected 
duration of the BH-NS merger events is
frequently somewhat longer than
the shortest timescale $T_{90}$ of a VSB (in our sample, the shortest VSB is 
5.3 ms, see Table~\ref{tab:expfits}; in the SWIFT data the shortest burst found, GRB 090515, had $T_{90}$ of 36 ms).
 
The Newtonian simulations of the BH-NS merger \citep{Lee1999}
were run for the mass ratio $q=M_{\rm NS}/M_{\rm BH}=0.1-1.0$, and showed that after about 30 ms the neutron star was disrupted and formed a torus of a mass of few tenths of the solar mass. The accretion rate at that moment was on the order of 2-6 solar masses per second, and therefore the expected lifetime of the torus feeding the black hole in during the GRB event was $\tau_{\rm accr} = M_{\rm disc}/\dot M = 40-60$ ms.
The relativistic simulations of \citet{Setia2006}, performed for a range of black hole spin parameters, followed the evolution of such torus with a mass of 0.01-0.2 $M_{\odot}$,. They showed that in the end of the run, 
after 40-70 ms the mass of the torus is about half of the initial mass and the accretion rate drops to about 0.2-0.6  $M_{\odot}$/s. Therefore the lifetime of the central engine will be much longer than that.

In merger event the duration of the 
phenomenon in the case of neutron star - black hole merger is fixed by the 
distance between 
the two at the phase of the neutron star tidal disruption and later by the 
viscosity operating 
within the accretion disk. 

In numerical simulation of \citet{Shibata2008} using $\Gamma$-law equation of state with 
$\Gamma=2$ the disruption of the neutron star happened close to the innermost 
stable circular orbit. The disruption process lasts for about 5 ms and 
after that time 
the unswallowed remnants of the star form a torus of the size of about 50 km. The 
further stage of the accretion onto a black hole lasts about 100 ms for a 3 $M_{\odot}$ 
black hole, according to \citet{Janiuk2004}
but those computations adopted 50 $R_S$ for the size of the 
outer edge of the disk.  If those computations are rescaled to 6 $R_S$ the 
duration of the accretion phase may possibly be considerably shorter, 
consistent with the
decay timescale of the shortest burst.

Short duration of the burst would then be provided by a small value of the black hole 
mass 
in the binary - larger black hole than 3 $M_{\odot}$ would give too long burst duration.
Thus in a natural way VSB would be connected with small mass black holes, would be 
fainter in the absolute luminosity and thus statistically less distant than 
short bursts.
This last effect might explain the lack of cosmological effects in their V/Vmax 
distribution. 

On the other hand, computations by \citet{Metzger2008} suggest
that the duration of the NS-BH merger is longer if the disk mass (and in 
consequence, the total luminosity) is smaller, and for small disk 
masses (0.03 $M_{\odot}$)
the event lasts above 0.1 s, up to 1 s. If so, the VSB, likely to be intrinsically 
faint,  are not consistent
with VSB properties.

Another merger scenario is possible if two neutron stars merge, and the total 
mass of the system is small. In this case it is probable that a differentially 
rotating hypermassive neutron star forms and initially is a stable 
configuration against the gravitational collapse. Later a delayed gravitational 
collapse is induced by the energy dissipation through gravitational waves, e.g. \citet{Shibata2006}; see also \citet{Oech2007}. 
The collapse again leads to formation of hot, 
magnetized torus surrounding a rotating BH with the outer radius of high density 
region much smaller than 20 km, and in this case the expected lifetime of the 
torus in only $\sim 10$ ms. The detailed predictions likely depend on the 
assumptions about the magnetic field \citep[see e.g.][]{Liu2008} but in 
general the agreement of the predicted timescales and observed 
properties of VSB are in this case satisfactory.

\subsubsection{evaporation of primordial black holes}

The primordial black holes discussed here as a possible origin of
the very short bursts should have masses in the range of $10^{9}-10^{14}$ g,
depending on time of their evaporation.
The constraints for the abundance of the holes contribution to the dark matter density 
in the Universe, depending on their masses was
recently discussed by \citet{lacki2009} in a broad range of masses. The observed
gamma-ray background limits the contribution of PBH to the critical density of the Universe 
to $\Omega_{PBH} < 10^{-8}$, which 
was already set by \citet{page76}, and similar results are
obtained nowadays \citep{Carr2010}. 
The limits for the somewhat larger PBH masses, obtained 
from the Milky Way neutrino, electron-positron and 
gamma ray backgrounds are constraining the total abundance of PBHs in the Universe 
and their contribution to the dark matter density, 
although the limits are less stringent \citep[$\Omega_{PBH} < 10^{-4}$ for 
PBH masses between $2 \times 10^{15}$ g to 1000 $M_{\odot}$]{lacki2009}. 
Abramowicz et al. (2009) studied the possible observational 
signatures of PBHs in the mass range of $10^{16}-10^{26}$ g via their collisions
with stars in our Galaxy, and found that the X-ray signals produced 
in such events would be too weak or too rare to be observable.

The indentification of the VSB with the evaporating black holes does not violate those
constraints. If the signal is indeed due to evaporating black holes, the 
typical distance to the source can be estimated in the following way.
The initial mass of the black hole evaporating now is of the 
order of $5 \times 10^{14}$ 
g, depending on the details of the evaporation efficiency at the later (hadron)
stages. Nevertheless, most of the mass disappears during the initial slow evaporation. 
The mass of the black hole at the final stages, 0.1 s before the hole 
disappearance, is only about   $6 \times 10^8$ g. If its energy is finally 
fully converted to photons and they reach BATSE detectors as $\sim 100 $ keV
the registration of a thousand photons would place the source at a distance of 
$\sim 2$ pc. Typical distance estimates were a factor of 100 
larger \citep[see e.g.][]{Cline2005}
since they were based on the initial PBH mass of $5 \times 10^{14}$ g. 
The numbers of VSB in the BATSE catalog,
roughly 10 per year, thus imply the average rate of 0.3 gamma ray 
flashes per cubic pc.
We assume an initial steep power law distribution of the black hole masses 
(with a power law index larger than 2),
and we take into account the mass evolution in time due to evaporation. The
number of observed flashes is obtained from integration over the mass distribution 
from the minimum mass evaporating now ($6 \times 10^8$ g at the final 0.1 s)
to the mass which will evaporate after a year ($4 \times 10^{11}$ g). Comparing this
theoretical rate with the observed number of bursts we derive the normalization 
constant of the mass distribution. Integrating over the whole mass distribution, 
we obtain mass density of PBH equal $7 \times 10^{-39}$ g cm$^3$, i.e. 
$\Omega_{PBH} \approx 10^{-9}$.

One of the  expected properties of the black hole evaporation is the uniformity of the
phenomenon. Black hole looses its angular momentum during the initial slow stage of the 
evaporation (before the loss of 10 per cent of the mass). Thus the only parameter
describing the evaporating black hole is its mass. If the number of particles 
released in
a considered time unit is large, every event should look identically.  The details of the 
emission process and the expected final radiation spectrum from the fireball-type last 
stages have been considered in a number of papers 
\citep[for the review see e.g.][]{ Bugaev2009}. Although the predictions 
differ with respect to the energy peak of the spectrum as well as the spectral slope
in the gamma-ray range, none of the papers predict any departure from isotropy of the 
emission, and consequently, from the identical look of all events.

Our sample of VSB shows a variety of shapes, thus the events are not identical. This 
means that either the primordial black hole evaporation origin is ruled out, or the
evaporation process is in need of a considerable revision.

Such a revision is likely to come. As argued by \citet{MacGibbon2008}, the interaction 
between the particles emitted by evaporating black hole is probably not as strong, 
as considered in a number of previous papers 
\citep[e.g.][]{Kapusta2001,Dag2002}; 
see \citet{Bugaev2008} for a review. If the particles
finally escaping the gravity field of evaporating black hole move radially, 
since only almost purely radial velocity allows to leave the black hole vicinity, 
and the evaporation 
process proceeds through quark jets, there is indeed no thermalization involved at the 
beginning of the process and local fluctuations (i.e. in emission in various directions) 
may be large. The total energy of a VSB registered by an instrument 
(BATSE, in our case) is 
less than 1 erg per event, and thus all the registered emission of a given burst 
is likely to come from a single jet-like event close to a black hole horizon. If this is the
case, then the events do not have to be exactly identical, prone to the initial quark ejection
as well as tiny differences with respect to the inclination of a specific jet. However,
they nevertheless should be short and hard, as already argued by \citet{Cline1992}, 
thus consistent with the VSB properties.  

We fitted to the composite VSB data 
the time profile of the optically thin fireball emitted in the final stage
of the black hole evaporation. The rise timescale is in this case given by the 
analytic function, $L(t)\propto t^{-2/3}$, while the normalization is fitted to the 
gamma ray 
background level. Such a model well represented the shape of the composite profile. 
After the maximum, the fireball will have different time 
characteristics, since we can have the hadronization with jet production, direct 
gamma and graviton production, as well as a direct lepton production, and large
fluctuations in this process may explain the shape variety. This the black hole evaporation model 
is also consistent with the data for VSB.

The tentative determination of the redshift for three VSB detected by
 SWIFT does not
automatically rule out the evaporation scenario, because the connection with the candidate galaxies
is only claimed by the proximity of the objects on the sky and may be a 
projection effect.
However, a long X-ray fading tail of the GRB 050509B discussed in Sect.~\ref{sect:swift}  
may indeed be difficult to explain within
the evaporation scenario.

\section{Conclusions}
\label{sec:concl}

We studied the sample of the BATSE gamma ray bursts, limited to the special group of 
VSB
($T_{90}<0.1$ s). These bursts seem to originate in a special region in our Galaxy, 
however we found that their arrival is statistically consistent with being uniform in 
time.
The time profiles of these bursts are asymmetric, with one or two peaks. We found no 
correlation between the burst asymmetry and its hardness, as well as no clear evidence 
for the time lags between the hard and soft photons.

We found that the burst properties are most likely to be consistent with two 
origin mechanisms. The first plausible mechanism is the
binary NS-NS merger with low total masses passing through a phase of hypermassive neutron star.
The second one is the evaporation of a primordial black hole in the scenario 
of no photosphere formation.

\medskip

{\it Acknowledgements.} 
This work
was supported in part by grants NN 203 380136 and NN 203 512638.
Part of this work was also supported 
by COST Action MP0905 'Black Holes in a violent Universe'.
This research has made 
use of BATSE data obtained from the High Energy Astrophysics Science Archive 
Research Center (HEASARC), provided by NASA's Goddard Space Flight Center.


\bibliographystyle{elsarticle-harv}

\end{document}